# Experimental demonstration of acoustic wave induced magnetization switching of dipole coupled magnetostrictive nanomagnets for ultralow power computing

Vimal Sampath[1], Noel D'Souza[1], Gary M. Atkinson[2], Supriyo Bandyopadhyay[2], and Jayasimha Atulasimha[1, 2] *

[1]Dept. of Mechanical and Nuclear Engineering, Virginia Commonwealth Univ., Richmond, VA 23284, USA

[2]Dept. of Electrical and Computer Engineering, Virginia Commonwealth Univ., Richmond, VA 23284, USA

* Corresponding author: jatulasimha@vcu.edu

We report nanomagnetic switching with Acoustic Waves (AW) launched from interdigitated electrodes that modulate the stress anisotropy of elliptical cobalt nanoscale magnetostrictive magnets (340 nm x 270 nm x 12 nm) delineated on 128° Y-cut lithium niobate. The dipole-coupled nanomagnet pairs are in a single-domain state and are initially magnetized along the major axis of the ellipse, with their magnetizations parallel to each other. The magnetizations of nanomagnets having lower shape anisotropy are reversed upon acoustic wave propagation. Thereafter, the magnetization of these nanomagnets remains in the "reversed" state and demonstrate non-volatility. This executes a 'NOT' operation. This proof of acoustic wave induced magnetic state reversal in dipole-coupled nanomagnets implementing a 'NOT' gate operation could potentially lead to the development of extremely energy-efficient nanomagnetic logic. Furthermore, fabrication complexity is reduced immensely due to the absence of individual contacts to the nanomagnets, leading to lower energy dissipation.

KEYWORDS: Acoustic Waves, magnetization switching, piezoelectric, magnetostrictive, multiferroic, lithium niobate

Nanomagnetic logic and memory devices have the benefit of non-volatility and energy efficiency [1, 2]. However, the challenge is to find an energy-efficient paradigm to switch the magnetic state of nanomagnets while dissipating very little energy in the process. The various ways to induce magnetization switching include the use of electric current-generated magnetic field [3], spin transfer torque [4], current-driven domain wall motion [5], current-induced spin orbit torque or spin hall effect [6, 7], or through strain generated by applying an electrical voltage to a multiferroic nanomagnet [8-14]. The energy dissipated in magnetization switching by applying strain in two phase elastically coupled multiferroic nanomagnets can be as low as 0.6 atto-Joules [15, 16], making it a promising switching mechanism. This strain is generated by application of an electrostatic potential to the piezoelectric layer and transferred to the magnetostrictive nanomagnet elastically coupled to it to implement an energy efficient method of magnetic state-switching as discussed in [13, 17].

Strain can be generated in a two-phase multiferroic nanomagnet by direct application of a voltage across the piezoelectric layer [11, 12] using contact pads. However, for some applications, such as Bennett clocking of *pipelined* dipole coupled logic [18], it would be necessary to apply strain sequentially to successive nanomagnets in an array. This would be lithographically challenging in an array of nanomagnets of feature size ~100 nm and pitch 300-500 nm as it requires individual contact pads around each nanomagnet. An acoustic wave launched in the direction of the array with the appropriate wavelength, on the other hand, will sequentially generate local stress underneath each nanomagnet (when a crest of the wave reaches a nanomagnet), thereby effectively carrying out Bennett clocking of pipelined dipole coupled logic *without* lithographic contacts to individual nanomagnets. This affords the best of both worlds - pipelined nonmagnetic logic for high speed computing and minimal lithography for high yield and low cost. The acoustic wave required must have a phase velocity that is slow enough so that the stress dwells long enough for the reversal to occur. Therefore, there is a need to reduce the acoustic wave velocity and this has been theoretically shown to be as low as 470 m/s for *X*-cut PMN-33%PT poled along [111]$_c$ [19]. In this paper, we demonstrate a simple proof of concept of achieving a NOT gate using dipole coupled multiferroic nanomagnets.

The use of Surface Acoustic Wave (SAW) to lower energy dissipation in switching of nanomagnets with spin transfer torque has been studied theoretically [20]. The periodic switching of magnetization between hard and easy axis of 40 μm × 10 μm × 10 nm Co bars sputtered on GaAs [21] and Ni films [22] was experimentally shown while the excitation of spin wave modes in a (Ga, Mn) As layer by a pico-second strain pulse has also been demonstrated [23]. On in-plane magnetized systems, SAWs have been used to drive ferromagnetic resonance in thin Ni films [24, 25]. Recent theoretical work discusses the possibility of complete reversal of magnetization with acoustic pulses [26, 27]. The effect of acoustic waves on magnetization switching in single domain isolated nanomagnets to a non-volatile vortex state has also been demonstrated [28]. The pre-stress state, which is a single domain state, goes into a vortex state upon application of the acoustic waves, and remains in the "vortex" state even after the waves have propagated through and there is no longer any strain in the nanomagnets. The vortex state is therefore non-volatile.

In this paper, we demonstrate, for the first time, acoustic wave-based magnetization reversal of single domain, dipole coupled, elliptical cobalt nanomagnets that implements a Boolean 'NOT' gate. In the pre-stress state, the magnetization of the nanomagnets are 'initialized' by a strong external magnetic field (FIG. 1a) and are single-domain in nature. Upon wave propagation (FIG. 1b), the magnetization of the nanomagnets having lower shape anisotropy switches and 'reverses' due to the dipole interaction with the neighboring nanomagnet having higher shape anisotropy (the stress anisotropy energy is unable to overcome this high shape anisotropy). This 'reversed' magnetization state persists even after the acoustic wave has propagated and is no longer straining these nanomagnets. Magnetic Force Microscopy (MFM) is used to characterize the nanomagnets' magnetic state before and after the acoustic wave clocking cycle is applied.

The wavelength of the acoustic wave is set to be 800 μm. This is to ensure that the wavelength is large enough that the strain across the nanomagnets is uniform. The Interdigitated transducers (IDTs) are a comb-like arrangement of rectangular aluminum bars of thickness 300 um and gap of 100 μm. The pitch of the IDTs is 400 μm which is exactly half the value of the intended acoustic wave wavelength.

The transmitter response function is μ(f), which is a function of the frequency of applied voltage, f. This transmitter response function is, in turn, a product of the single tap response function $μ_s(f, η)$ and Array factor, H(f):

$$μ(f) = μ_s(f, η)\, H(f),  \quad (1)$$

$$μ_s(f, η) = μ_s(f_0, η) \sin(πf/2f_0),  \quad (2)$$

$$H(f) = N \sin Nπ\,[(f-f_0/f_0)] / Nπ\,[(f-f_0/f_0)],  \quad (3)$$

The single tap response function varies with frequency, f, and the metallization ratio, η. For an applied frequency of $f = f_0 = 5$ MHz and a metallization ratio of 0.75, $μ_s(f, η) = 0.9K^2$ where $K^2 = 0.056$ [29]. When the frequency of applied voltage is equal to the characteristic frequency of the IDTs, the array factor is equal to the pairs of electrodes in the transmitter IDT, N which is 40 in the current design. Therefore, the transmitter response function, μ(f), is calculated to be 2.016.

The IDTs are fabricated using conventional photolithography and wet etching processes. Two sets of IDTs are fabricated (as shown in FIGS. 1c, d). One set is used to launch the acoustic waves and the other is used to sense the propagated acoustic waves. For the purpose of these experiments, the receiver transducer is redundant and is only used to check electrical connections and confirm wave propagation (see Supplementary Section B).

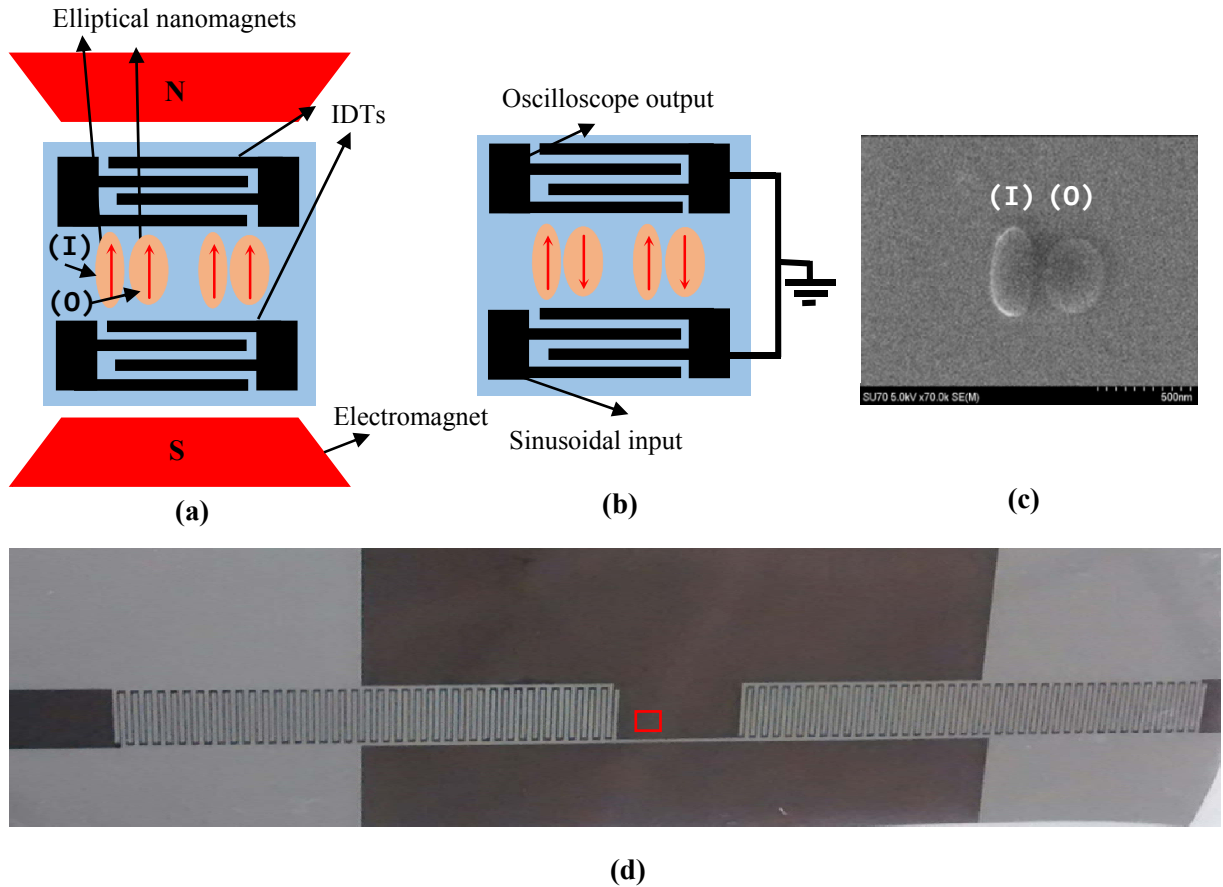

**FIG 1:** (a) Schematic of experimental set-up with initial application of an external magnetic field on the nanomagnets. The red arrows indicate the direction of the magnetization state of the nanomagnets. (b) Upon acoustic wave propagation, a mechanical strain is generated and transferred to the nanomagnets which reverses the magnetization of the lower shape anisotropy nanomagnets (O). (c) SEM micrograph of the dipole coupled nanomagnets with nominal dimensions of 400 nm × 150 nm × 12 nm (I) and 340 nm × 270 nm × 12 nm (O). (d) Optical image of the lithium niobate substrate with the fabricated IDTs. The red rectangle highlights the region containing the nanomagnets in the delay line.

Dipole-coupled elliptical Co nanomagnets of dimensions ~ 400 nm x 150 nm x 12 nm termed an input nanomagnet (I) and 340 nm × 270 nm × 12 nm termed as output nanomagnet (O) were fabricated on a 128° Y-cut lithium niobate substrate. Prior to the nanofabrication process, the substrate was spin-coated with a bi-layer PMMA e-beam resist of different molecular weights in order to obtain a greater degree of undercut: PMMA-495 (diluted 4% V/V in Anisole) followed by PMMA-950 (diluted 4% V/V in Anisole) at a spin rate of 2000 rpm. The resists were baked at 90° C for 5 minutes. Next, electron-beam lithography is performed using a Hitachi SU-70 Scanning Electron Microscope (at an accelerating voltage of 30 kV and 60 pA beam current) with a Nabity NPGS lithography system. Subsequently, the resists were developed in MIBK:IPA (1:3) for 270 seconds followed by a cold IPA rinse. For nanomagnet delineation, a 5 nm thick Ti adhesion layer was first deposited using e-beam evaporation at a base pressure of ~2 × $10^{-7}$ Torr, followed by the deposition of 10 nm of Co. The liftoff was carried out using Remover PG solution.

The acoustic waves are excited and detected by aluminum Interdigital transducers (IDTs) fabricated on a piezoelectric lithium niobate substrate. The acoustic waves cause a Rayleigh mode of displacement on the surface of Lithium niobate in the delay line between the transmitter and receiver IDTs [29]. We note that while the mathematical treatment used is that for surface acoustic wave (SAW) propagation, we term these waves as "acoustic waves" as the penetration depth which is of the order of a wavelength is comparable to the thickness of the substrates. Hence, we do not term it a surface acoustic wave (SAW).

Elliptical nanomagnets are delineated in the delay line, as shown in FIGS. 1d and 1e. The pairs of nanomagnets of nominal dimensions (340 nm × 270 nm × 12 nm) and (400 nm × 150 nm × 12 nm) are initially magnetized along the major axis with a large external magnetic field of ~0.2 Tesla (FIG. 1a) and characterized by MFM (FIG. 3a). The magnetization orientation (single-domain) of these nanomagnets is found to be along the major axis, as expected.

Mechanical strain is applied by applying a sinusoidal voltage of 50 $V_{p-p}$ between the IDTs (FIG. 1d) at a characteristic frequency. The relationship between electrostatic potential, ϕ, and applied sinusoidal voltage, V, is

$$\phi = \mu(f) \cdot V, \qquad (4)$$

Here $\mu(f)$ is the transmitter response function. Using the earlier calculated value of $\mu(f)$ =2.016, electrostatic potential, ϕ, in the delay line of Lithium niobate is 100.8 V. The particle displacement on a lithium niobate substrate is known to be 0.18 nm per volt of electrostatic potential [26]. Thus the maximum strain generated by this acoustic wave over a length of 340 nm is calculated to be 142.5 ppm, as explained in Supplementary Section B.

The displacement wave can be expressed as 18.1 sin(2πx/λ) nm, as shown in FIG. 2. Here, λ is the wavelength of the acoustic wave which is 800 μm. To calculate maximum strain over a length of 340 nm, we need to calculate the displacement at x = 170 nm and x = -170 nm. This is because the strain is maximum around x = 0. The displacement at x = 170 nm is 0.0242 nm and at x = -170 nm is -0.0242 nm. Therefore, the total change is length is 0.0483 nm, and the strain is 0.0483/340 = 142.16 ppm. If we assume the Young's modulus of Co nanomagnets to be equal to the bulk Young's modulus of 209 GPa, the stress is 29.71 MPa.

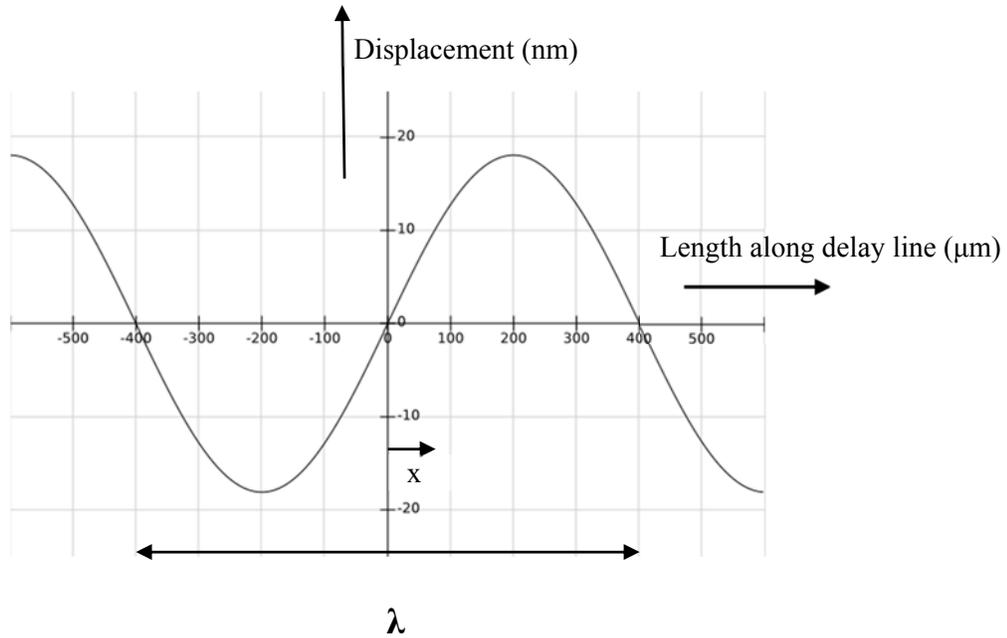

FIG. 2. The picture showing the displacement wave of points in the delay line of Lithium niobate

Assuming that this maximum strain is completely transferred to the multiferroic nanomagnet, the maximum stress applied to the elliptical Co nanomagnets is 30 MPa. During this stress (voltage) application, each nanomagnet experiences cycles of tensile and compressive stress (± 30 MPa) along its major axis. We note that multiple cycles of acoustic waves pass through the nanomagnets, and not just a singular pulse. Since Cobalt has negative magnetostriction, the tensile stress on the lower shape anisotropic nanomagnet (340 nm × 270 nm × 12 nm) results in magnetization rotation towards the minor axis. This is because stress anisotropy shifts the potential energy minimum to an orientation that is perpendicular to the stress axis. The higher shape anisotropic nanomagnets (400 nm × 150 nm × 12 nm) have a greater shape anisotropy energy barrier which the stress anisotropy is unable to beat. After the acoustic wave propagation through the nanomagnets, the magnetization of lower shape anisotropic nanomagnets has an equal probability of either returning to its original orientation or switching its orientation by 180°, since the potential energy landscape reverts to its original state when the stress is withdrawn. This assumption is based on single-domain magnetization switching. As seen in the MFM image of FIG. 3b, after acoustic wave propagation, the magnetization state of the nanomagnets having lower shape anisotropy energy barrier (O) switches by 180° due to the dipole interaction with the higher shape anisotropic nanomagnet (I). Hence, the magnetization state of the lower shape anisotropic magnets, dipole coupled with the higher shape anisotropic nanomagnet, has been reversed with application of acoustic wave. This magnetization state remains in this 'reversed' state even upon removal of the acoustic wave.

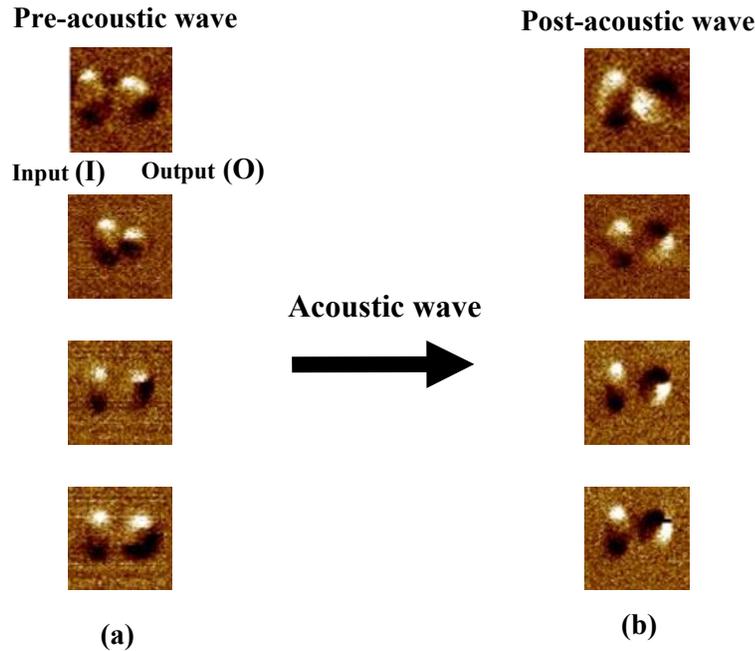

FIG. 3. MFM images of six distinct nanomagnet pairs. (a) Pre-stress (pre-acoustic wave) magnetization state. (b) Post-stress (post-acoustic wave) magnetization state. The lower shape anisotropic nanomagnets (O) clearly show a magnetization rotation of 180°.

In summary, we have shown that acoustic waves, generated by IDTs fabricated on a piezoelectric lithium niobate substrate, can be utilized to induce 180° magnetization switching in dipole coupled elliptical Co nanomagnets. The magnetization switches from its initial single-domain 'up' state to a single-domain 'down' state after tensile/compressive stress cycles propagate through the nanomagnets. The switched state is stable and non-volatile. Furthermore, the acoustic wave energy amortized over all the nanomagnet pairs that can potentially fit in the delay line (as discussed in supplementary section C) could result in energy dissipation of the order of tens of attojoules per nanomagnet per clock cycle. These results show the feasibility of an extremely energy efficient acoustically clocked Boolean NOT gate.

The authors gratefully acknowledge the US National Science Foundation for financial support, under the SHF-Small Grant CCF-1216614, CAREER Grant CCF-1253370. We also acknowledge the assistance provided by David Gawalt, Joshua Smak and Catherine Fraher who were involved in the initial design and fabrication of the IDTs as part of their senior design project, as well as Shopan Hafiz and Prof. Umit Ozgur for assisting with the initial operation of the pulse generator. We also acknowledge the use of the CNST facility at NIST, Gaithersburg, Maryland, USA for some of the nanofabrication work.

# Experimental demonstration of acoustic wave induced magnetization switching of dipole coupled magnetostrictive nanomagnets for ultralow power computing

Vimal Sampath[1], Noel D'Souza[1], Gary M. Atkinson[2], Supriyo Bandyopadhyay[2] and Jayasimha Atulasimha[1,2] *

[1]Department of Mechanical and Nuclear Engineering

[2]Department of Electrical and Computer Engineering

Virginia Commonwealth University, Richmond, VA 23284, USA.

* Corresponding author: jatulasimha@vcu.edu

**Supplementary Section A: Magnetic Force Microscopy (MFM) scans to demonstrate repeatability**

The images shown in top two rows in FIG. S1 are MFM phase images of the two bottom magnets shown in FIG. 3 of the main paper. Here we confirm that on application of a magnetic field they can reset to the initial state. The next two sets of images (row 3 and 4) show different nanomagnets where we not only show the "reset" after the switching but also demonstrate that the entire switching sequence is repeatable.

To establish repeatability and 'reset' the magnetization of the nanomagnets to the original pre-stress magnetization state, a large external magnetic field of 0.2 Tesla is applied along the major axes of the nanomagnets as shown in FIG. 1a of main paper. The MFM images of exactly the same nanomagnets after this 'reset' step are shown in FIG. S1c. The images clearly show that the single domain pre-stress state (before the acoustic wave is applied) of the magnetization has been restored since the images in FIG. S1c are nearly identical to those in FIG. S1a.

Acoustic wave (AW) is again applied as shown in FIG. 1b (of main paper). The magnetization of the bottom nanomagnets are again reversed. The resulting MFM images are shown in FIG. S1d. The images clearly show that the magnetization again goes into same reversed state since the images in FIG. S1d are nearly identical to those in FIG. S1b.

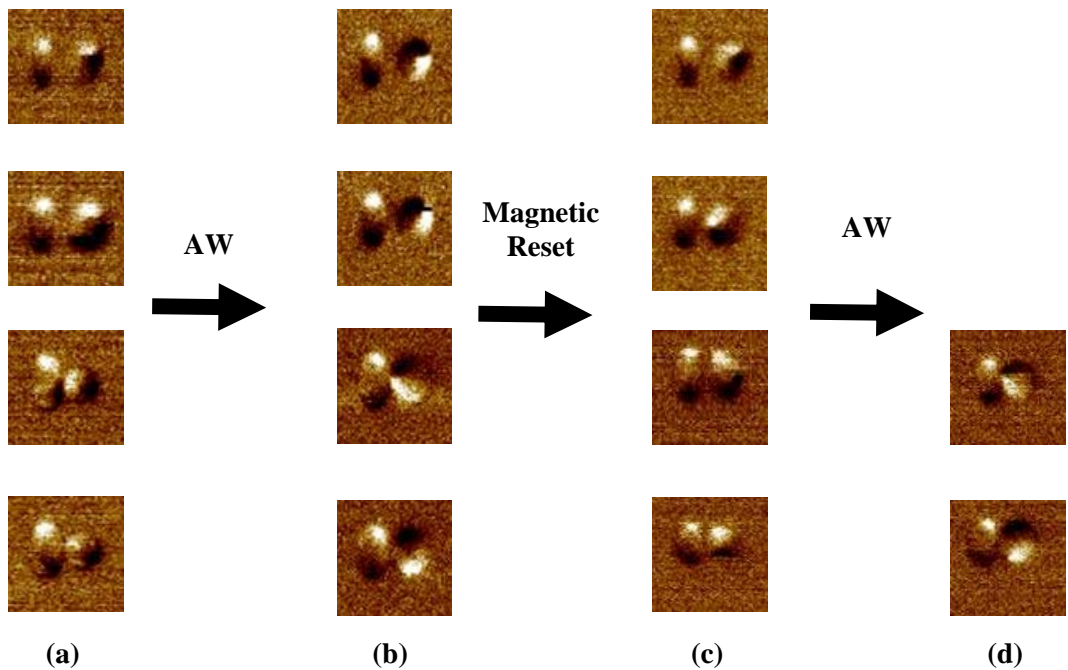

FIG. S1: MFM images of 4 dipole coupled nanomagnet pairs in the: (a) Pre-stress state before AW application. (b) Post-AW images. (c) The third column shows MFM images taken after magnetization of the nanomagnet pair is reset with a magnetic field. (d) This is the post AW images when AW is applied to nanomagnets.

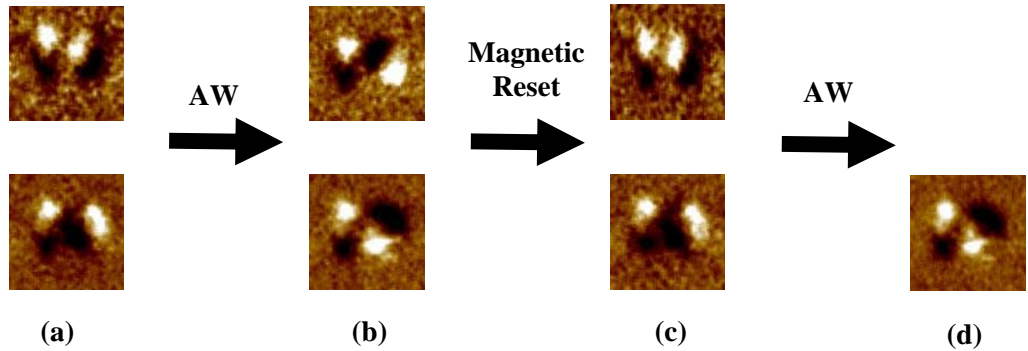

FIG. S2: Two sets of dipole coupled magnets where applied AW wavelength is doubled.

FIG. S2 shows MFM images of two different pairs of nanomagnets which have the same shape as magnets in S1. Unlike the acoustic wave applied to magnets shown in S1 where the wavelength was 800 μm, the applied acoustic wave wavelength was 1600 μm. FIG. S2 shows the same kind of repeatability and MFM scans as FIG. S1. Since, the wave dwells on the nanomagnet for sufficient time in the former case, additional dwell time in the latter case causes no significant difference in switching behavior.

**Supplementary Section B: Electrical characterization of IDTs**

The FIGS. S3 (a) and (b) show the oscilloscope image when sinusoidal and square wave voltage respectively are applied to the IDTs fabricated on glass. There is no transmission as expected for a glass substrate.

The FIGS. S4 (a) and (b) show the oscilloscope images of sinusoidal and square wave voltage respectively are applied to the IDTs fabricated on Lithium niobate. It can be clearly seen that the receiver IDTs on Lithium niobate show sinusoidal electrical signal and the peak-peak voltage of this signal is almost equal that applied to the transmitted IDTs thus proving no significant loss of amplitude in the delay line. We also note that is the output voltage in FIG. S4 (b) is sinusoidal, even though applied voltage is a square wave.

Thus, the acoustic wave device acts as a filter when the applied voltage is of the frequency of the characteristic frequency of the device, as discussed in literature[1].

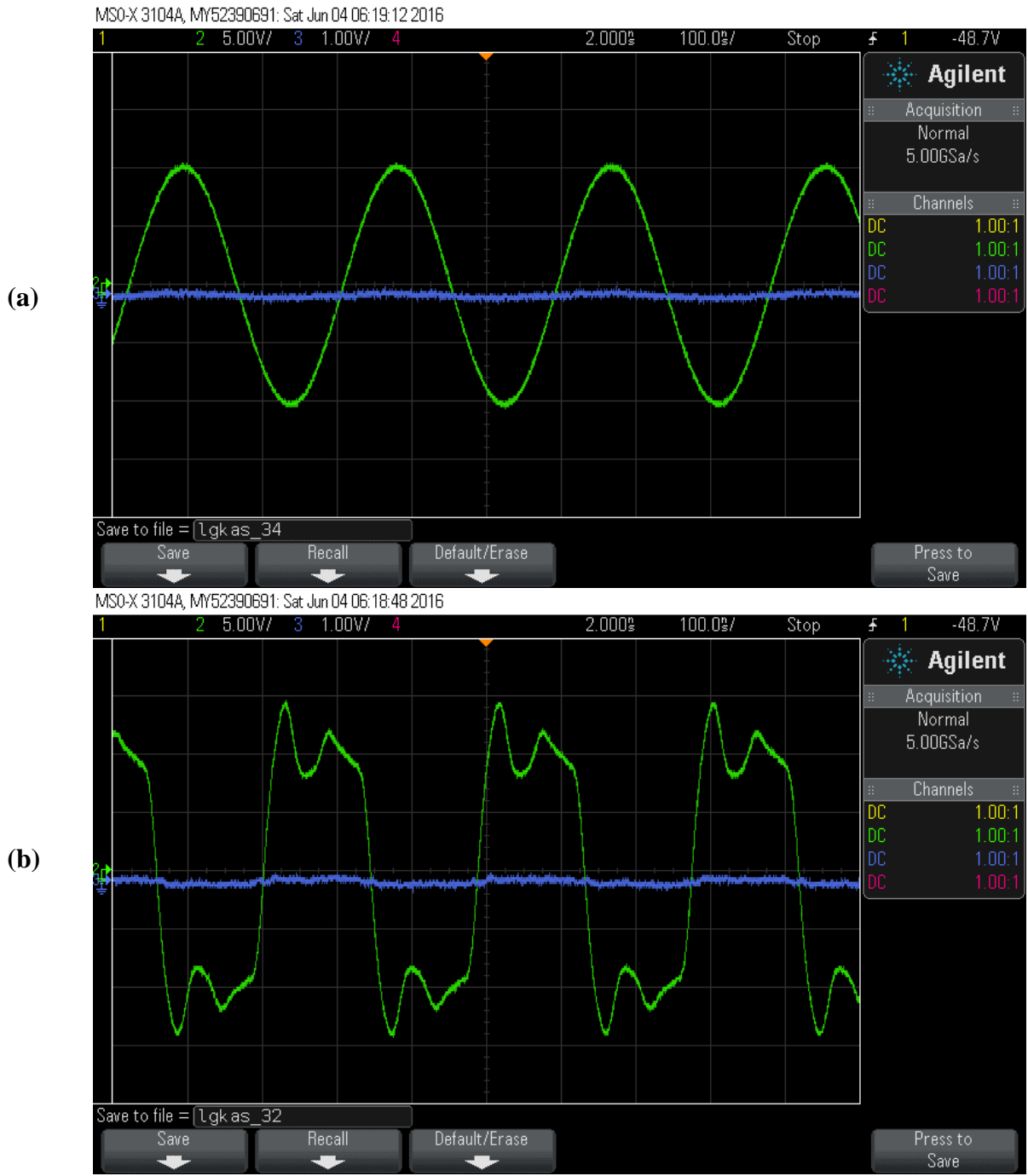

FIG. S3. (a) Sinusoidal voltage applied to input IDT on Glass. (b) Square wave applied to input IDT on Glass. The input is shown in green. The output from receiver IDT is in blue. Shows NO transmission as expected.

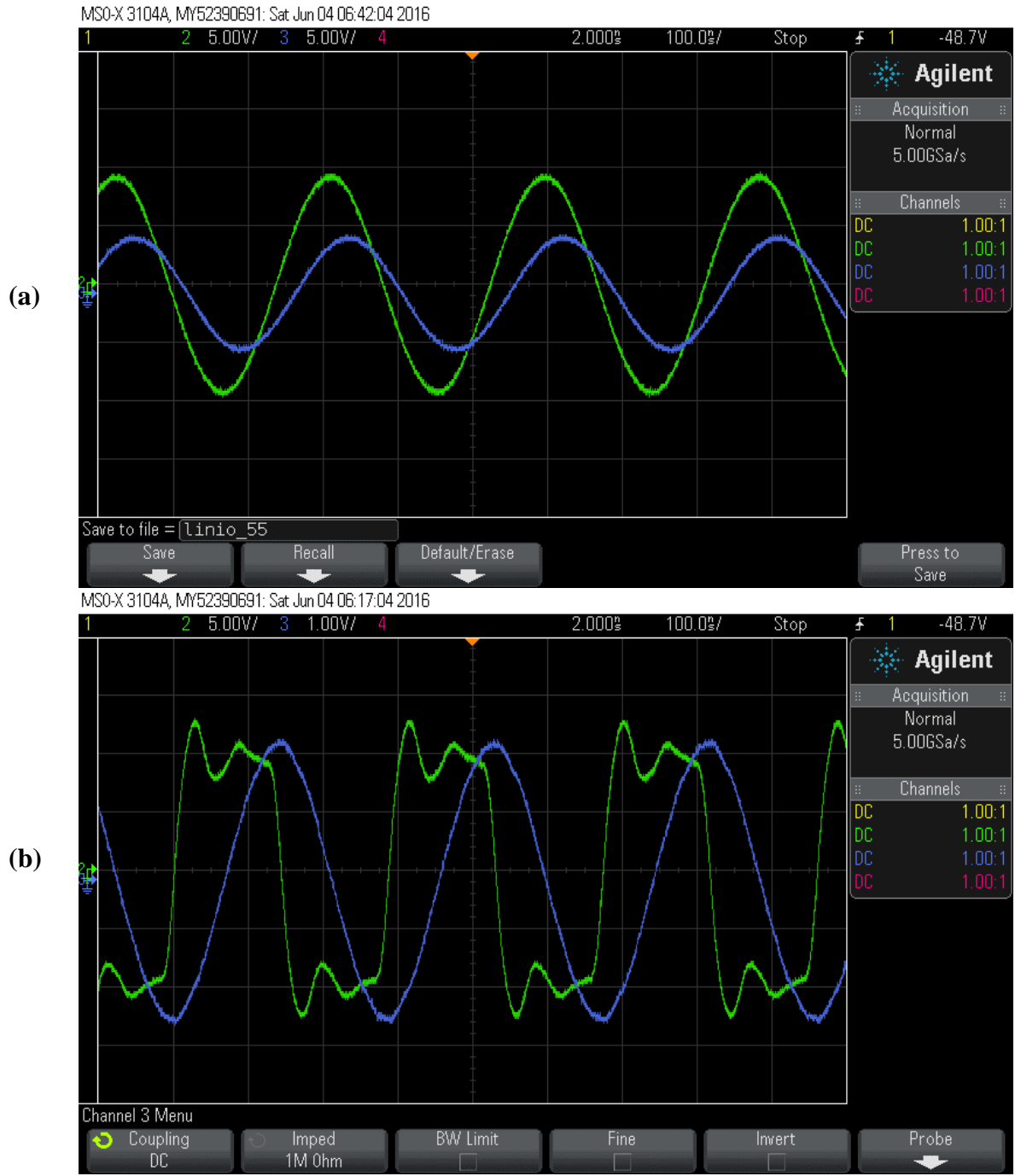

FIG. S4. (a) Sinusoidal voltage applied to input IDT on Lithium Niobate. (b) Square wave applied to input IDT on Lithium Niobate. The input is shown in green. The output from receiver IDT is in blue. Shows large transmission

**Supplementary Section C: Energy dissipation in nanomagnets due to the acoustic wave**

The total power per beam width generated by an acoustic wave when a voltage is applied to the IDTs is given by[1]

$$\frac{P}{W} = \frac{1}{2}\phi^2 \left(\frac{y_0}{\lambda}\right) \qquad (9)$$

where $\phi$ is the surface potential (~100.8 V to generate a stress of 30 MPa), $y_0$ is the admittance of the lithium niobate substrate (0.21 × 10$^{-3}$ S), $W$ is the beam width of the IDTs, and $\lambda$ is the acoustic wave wavelength ($\lambda = 800\ \mu m$). Therefore, the total power density per unit beam width is calculated to be 1333.6 W/m. In order to estimate the energy dissipation per nanomagnet, we consider the nanomagnets used in the experiment having lateral dimensions of 340 nm × 270 nm and assume a 2-dimensional array of such magnets can be designed with a center-to-center separation of ~0.85 μm between the nanomagnets along the AW propagation direction and ~0.5 μm perpendicular to it. With negligible AW attenuation at low frequencies (less than 0.1% over a length of 1 cm, at 10 MHz[2]), it can be safely assumed that at a frequency of 5 MHz, the AW wave can clock a ~8 cm long chain of nanomagnets with minimal attenuation. Considering an IDT beam width of 1 cm, a single AW cycle can clock ~1.88 billion nanomagnets and the energy dissipated per nanomagnet for one AW cycle (tension and compression) of time period 200 ns is ~1.4 fJ. If the clocking frequency is increased a hundred times to ~500 MHz while the power is kept constant (as less stress over smaller time is needed if materials with large magneto-elastic coupling such as Terfenol-D are used) the energy dissipation can be decreased to a mere ~14 aJ per nanomagnet. This would make the AW based clocking extremely energy efficient without requiring lithographic contacts to each and every nanomagnet.